\documentclass[pre,showpacs,twocolumn]{revtex4}
\usepackage{graphicx}

\begin{document}

\title{Neural cryptography with feedback}
\date{26 November 2003}
\author{Andreas Ruttor}
\author{Wolfgang Kinzel}
\affiliation{Institut f\"ur Theoretische Physik, Universit\"at
  W\"urzburg, Am Hubland, 97074 W\"urzburg, Germany}
\author{Lanir Shacham}
\author{Ido Kanter}
\affiliation{Department of Physics, Bar Ilan University, Ramat Gan
  52900, Israel}
\begin{abstract}
  Neural cryptography is based on a competition between attractive and
  repulsive stochastic forces. A feedback mechanism is added to neural
  cryptography which increases the repulsive forces. Using numerical
  simulations and an analytic approach, the probability of a
  successful attack is calculated for different model parameters.
  Scaling laws are derived which show that feedback improves the
  security of the system. In addition, a network with feedback
  generates a pseudorandom bit sequence which can be used to encrypt
  and decrypt a secret message.
\end{abstract}
\maketitle

\section{Introduction}

Neural networks learn from examples. When a system of interacting
neurons adjusts its couplings to a set of externally produced
examples, this network is able to estimate the rule which produced the
examples. The properties of such networks have successfully been
investigated using models and methods of statistical
physics~\cite{Hertz:1991:ITN, Engel:2001:SML}.

Recently this research program has been extended to study the
properties of interacting networks~\cite{Metzler:2000:INN,
  Kinzel:2000:DIN}. Two networks which learn the examples produced by
their partner are able to synchronize. This means that after a
training period the two networks achieve identical time dependent
couplings (synaptic weights). Synchronization by mutual learning is a
phenomenon which has been applied to
cryptography~\cite{Kanter:2002:SEI, Kinzel:2003:DGI}.

To send a secret message over a public channel one needs a secret key,
either for encryption, decryption, or both. In 1976, Diffie and
Hellmann have shown how to generate a secret key over a public channel
without exchanging any secret message before. This method is based on
the fact that---up to now---no algorithm is known which finds the
discrete logarithm of large numbers by feasible computer
power~\cite{Stinson:1995:CTP}.

Recently it has been shown how to use synchronization of neural
networks to generate secret keys over public
channels~\cite{Kanter:2002:SEI}. This algorithm, called neural
cryptography, is not based on number theory but it contains a physical
mechanism: The competition between stochastic attractive and repulsive
forces. When this competition is carefully balanced, two partners $A$
and $B$ are able to synchronize whereas an attacking network $E$ has
only a very low probability to find the common state of the
communicating partners.

The security of neural cryptography is still being debated and
investigated~\cite{Klimov:2002:ANC, Kinzel:2002:INN, Kinzel:2002:NC,
  Mislovaty:2002:SKE, Kanter:2002:TNN}. In this paper we introduce a
mechanism which is based on the generation of inputs by feedback.
This feedback mechanism increases the repulsive forces between the
participating networks, and the amount of the feedback, the strength
of this force, is controlled by an additional parameter of our model.

A measure of the security of the system is the probability $P_E$ that
an attacking network is successful. We calculate $P_E$ obtained from
the best known attack~\cite{Klimov:2002:ANC} for different model
parameters and search for scaling properties of the synchronization
time as well as for the security measure. It turns out that feedback
improves the security significantly, but it also increases the effort
to find the common key. When this effort is kept constant, feedback
only yields a small improvement of security.

\section{Repulsive and attractive stochastic forces}

\begin{figure}
  \centering
  \includegraphics[width=8.6cm]{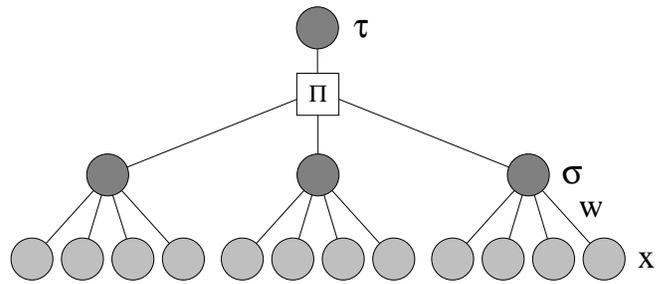}
  \caption{\label{fig:tpm} A tree parity machine with $K=3$ and
    $N=4$.}
\end{figure}

The mathematical model used in this paper is called a tree parity
machine (TPM), sketched in Fig.~\ref{fig:tpm}. It consists of $K$
different hidden units, each of them being a perceptron with an
$N$-dimensional weight vector $\mathbf{w}_k$. When a hidden unit $k$
receives an $N$-dimensional input vector $\mathbf{x}_k$ it produces
the output bit
\begin{equation}
  \sigma_k=\textrm{sgn}(\mathbf{w}_k \cdot \mathbf{x}_k)\,.
\end{equation}
The $K$ hidden units $\sigma_k$ define a common output bit $\tau$ of
the total network by
\begin{equation}
  \tau = \prod\limits_{k=1}^K \sigma_k\,.
\end{equation}

In this paper we consider binary input values $x_{k,j} \in \{-1,+1\}$
and discrete weights $w_{k,j} \in \{-L,-L+1,...,L-1,L\}$, where the
index $j$ denotes the $N$ components and $k$ the $K$ hidden units.

Each of the two communicating partners $A$ and $B$ has its own network
with an identical TPM architecture. Each partner selects random
initial weight vectors $\mathbf{w}_k^A(t=0)$ and
$\mathbf{w}_k^B(t=0)$.

Both of the networks are trained by their mutual output bits $\tau^A$
and $\tau^B$. At each training step, the two networks receive common
input vectors $\mathbf{x}_k$ and the corresponding output bit $\tau$
of its partner. We use the following learning rule.

\begin{enumerate}
  \renewcommand{\labelenumi}{(\arabic{enumi})}
  \renewcommand{\labelenumii}{(\roman{enumii})}
\item If the output bits are different, $\tau^A \neq \tau^B$, nothing
  is changed.
\item If $\tau^A = \tau^B \equiv \tau$ only the hidden units are
  trained which have an output bit identical to the common output,
  $\sigma_k^{A/B} = \tau^{A/B}$.
\item To adjust the weights we consider three different learning
  rules.
  \begin{enumerate}
  \item Anti-Hebbian learning
    \begin{equation}
      \label{eq:anti-hebb}
      \mathbf{w}_k^{+} = \mathbf{w}_k - \tau \mathbf{x}_k
      \Theta(\sigma_k \tau) \Theta(\tau^A \tau^B)\,.
    \end{equation}
  \item Hebbian learning
    \begin{equation}
      \mathbf{w}_k^{+} = \mathbf{w}_k + \tau \mathbf{x}_k
      \Theta(\sigma_k \tau) \Theta(\tau^A \tau^B)\,.
    \end{equation}
  \item Random walk
    \begin{equation}
      \mathbf{w}_k^{+} = \mathbf{w}_k + \mathbf{x}_k
      \Theta(\sigma_k \tau) \Theta(\tau^A \tau^B)\,.
    \end{equation}
  \end{enumerate}
  If any component $w_{k,j}$ moves out of the interval $-L,\dots,L$,
  it is replaced by $\mathrm{sgn}(w_{k,j}) L$.
\end{enumerate}

Note that for the last rule, the dynamics of each component is
identical to a random walk with reflecting boundaries. The only
difference to usual random walks is that the dynamics is controlled by
the $2 K$ global signals $\sigma_k^{A/B}$ which, in turn, are
determined by the ensemble of random walks. Two corresponding
components of the weights of $A$ and $B$ receive an identical input
$x_{k,j}$, hence they move into the same direction if the control
signal allows both of them the move. As soon as one of the two
corresponding components hits the boundary their mutual distance
decreases. This mechanism finally leads to complete synchronization,
$\mathbf{w}^A_k(t)=\mathbf{w}^B_k(t)$ for all $t \geq t_{sync}$.

On average, a common step leads to an attractive force between the
corresponding weight vectors. If, however, only the weight vector of
one of the two partners is changed the distance between corresponding
vectors increases, on average. This may be considered as a repulsive
force between the corresponding hidden units.

A learning step in at least one of the $K$ hidden units occurs if the
two output bits are identical, $\tau^A=\tau^B$. In this case, there
are three possibilities for a given pair of hidden units:
\begin{enumerate}
  \renewcommand{\labelenumi}{(\arabic{enumi})}
\item an attractive move for $\sigma_k^A=\sigma_k^B=\tau^{A/B}$;
\item a repulsive move for $\sigma_k^A\neq\sigma_k^B$;
\item and no move at all for $\sigma_k^A=\sigma_k^B\neq\tau^{A/B}$.
\end{enumerate}

We want to calculate the probabilities for repulsive and attractive
steps~\cite{Klimov:2002:ANC, Rosen-Zvi:2002:MLT}. The distance between
two hidden units can be defined by their mutual overlap
\begin{equation}
  \rho_k = \frac{\mathbf{w}^A_k \cdot
    \mathbf{w}^B_k}{\sqrt{\mathbf{w}^A_k \cdot \mathbf{w}^A_k} \;
    \sqrt{\mathbf{w}^B_k \cdot \mathbf{w}^B_k}}\,.
\end{equation}

The probability $\epsilon_k$ that a common randomly chosen input
$\mathbf{x}_k$ leads to a different output bit $\sigma_k^A \ne
\sigma_k^B$ of the hidden unit is given by~\cite{Engel:2001:SML}
\begin{equation}
  \epsilon_k = \frac{1}{\pi} \arccos \rho_k\,.
\end{equation}

The quantity $\epsilon_k$ is a measure of the distance between the
weight vectors of the corresponding hidden units. Since different
hidden units are independent, the values $\epsilon_k$ determine also
the conditional probability $P_r$ for a repulsive step between two
hidden units given identical output bits of the two TPMs. In the case
of identical distances, $\epsilon_k=\epsilon$, one finds for $K=3$
\begin{eqnarray}
  P_r &=& P(\sigma^A_k \neq \sigma_k^B | \tau^A = \tau^B) \nonumber \\
      &=& \frac{2 (1-\epsilon) \epsilon^2}{(1-\epsilon)^3 + 3
          (1-\epsilon) \epsilon^2}\,.
\end{eqnarray}

On the other side, an attacker $E$ may use the same algorithm as the
two partners $A$ and $B$. Obviously, it will move its weights only if
the output bits of the two partners are identical. In this case, a
repulsive step between $E$ and $A$ occurs with probability $P_r =
\epsilon$ where now $\epsilon$ is the distance between the hidden
units of $E$ and $A$.

Note that for both the partners and the attacker one has the important
property that the networks remain identical after synchronization.
When one has achieved $\epsilon=0$ at some time step, the distance
remains zero forever, according to the previous equations for $P_r$.
However, although the attacker uses the same algorithm as the two
partners, there is an important difference: $E$ can only listen but it
cannot influence $A$ or $B$. This fact leads to the difference in the
probabilities of repulsive steps; the attacker has always more
repulsive steps than the two partners. For small distances $\epsilon
\ll 1$, the probability $P_r$ increases linear with the distance for
the attacker but quadratic for the two partners. This difference
between learning and listening leads to a tiny advantage of the
partners over an attacker. The subtle competition between repulsive
and attractive steps makes cryptography feasible.

On the other side, there is always a nonzero probability $P_E$ that an
attacker will synchronize, too~\cite{Mislovaty:2002:SKE}. For neural
cryptography, $P_E$ should be as small as possible. Therefore it is
useful to investigate synchronization for different models and to
calculate their properties as a function of the model parameters.

Here we investigate a mechanism which decreases $P_E$, namely we
include feedback in the neural networks. The input vectors
$\mathbf{x}_k$ are no longer common random numbers, but they are
produced by the bits of the corresponding hidden units. Therefore the
hidden units of the two partners no longer receive an identical input,
but two corresponding input vectors separate with the number of
training steps. To allow synchronization, one has to reset the two
inputs to common values after some time interval.

For nonzero distance $\epsilon > 0$, this feedback mechanism creates a
sort of noise and increases the number of repulsive steps. After
synchronization $\epsilon = 0$, feedback will produce only identical
input vectors and the networks move with zero distance
forever~\cite{comment}\nocite{Mislovaty:2003:PCC}.

Before we discuss synchronization and several attacking scenarios, we
consider the properties of the bit sequence generated by a TPM with
feedback.

\section{Bit Generator}

We consider a single TPM network with $K$ hidden units, as in the
preceding section. We start with $K$ random input vectors
$\mathbf{x}_k$. But now, for each hidden unit $k$ and for each time
step $t$, the input vector is shifted and the output bit $\sigma_k(t)$
is added to its first component~\cite{Eisenstein:1995:GPT}.
Simultaneously, the weight vector $\mathbf{w}_k$ is trained according
to the anti-Hebbian rule, Eq.~(\ref{eq:anti-hebb}). Consequently, the
bit sequence $\tau(t)$ generated by the TPM is given by the equation
\begin{equation}
  \tau(t)= \prod \limits_{k=1}^K \textrm{sgn} \left( \sum
    \limits_{j=1}^N w_{k,j}(t) \sigma_k(t-j) \right)\,.
\end{equation}

Similar bit generators were introduced in Ref.~\cite{Zhu:1998:APS} and
the statistical properties of their generated sequences were
investigated~\cite{Metzler:2001:GUT}. Here we study the corresponding
properties for our TPM with discrete weights.

The TPM network has $2^{KN}$ possible input and $(2L+1)^{KN}$ weight
vectors. Therefore our deterministic finite state machine can only
generate a periodic bit sequence whose length $l$ is limited by
$(4L+2)^{KN}$.

Our numerical simulations show that the average length $\langle l
\rangle$ of the period indeed increases exponentially fast with the
size $KN$ of the network, but it is much smaller than the upper bound.
For $K=3$ and $L>N$ we find $\langle l \rangle \propto (2.69)^{3N}$,
independent of the number $L$ of weight values.

The network takes some time before it generates the periodic part of
the sequence. We find that this transient time also scales
exponentially with the system size $KN$. This means that, for
sufficiently large values of $N$, say $N \geq 100$, any simulation of
the bit sequence remains in the transient part and will never enter
the cycle.

The bit sequence generated by a TPM with $K>2$ cannot be distinguished
from a random bit sequence. For $K=L=3$ we have numerically calculated
its entropy and found the value $\ln 2$ as expected from a truly
random bit sequence. In addition, we have performed several tests on
randomness as described by Knuth~\cite{Knuth:1981:SA}. We did not find
any correlations between consecutive bits; the bit sequence passed all
tests on randomness within strict confidence levels.

Although the bit sequence passed many known tests on random numbers we
know that it is generated by a neural network. Does this knowledge
help to estimate correlations of the sequence and to predict it? In
fact, for a sequence generated by a perceptron (TPM with $K=1$),
another perceptron trained on the sequence could achieve an overlap to
the generator~\cite{Metzler:2000:INN}.

Consider a bit sequence generated by a TPM with the anti-Hebbian rule.
Another TPM (the ``student'') is trained on this sequence using the
same rule. In addition, if the output bit disagrees with the
corresponding bit of the sequence, we use the geometric method of
Ref.~\cite{Klimov:2002:ANC} to perform a training step.

\begin{figure}
  \centering
  \includegraphics[width=8.6cm]{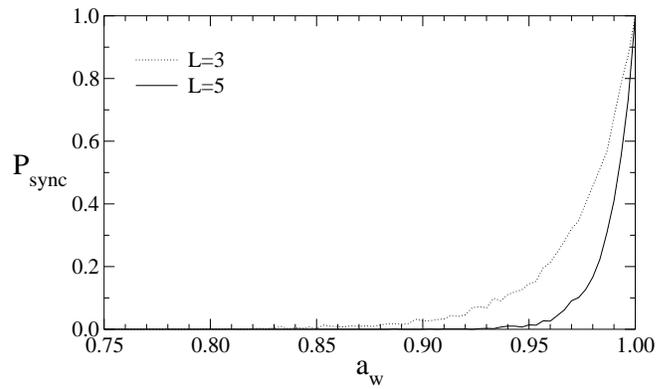}
  \caption{\label{fig:bg_pe} Probability $P_{sync}$ as a function of
    the fraction $a_w$ of initially known weights, calculated from
    1000 simulations with $K=3$ and $N=100$.}
\end{figure}

\begin{figure}
  \centering
  \includegraphics[width=8.6cm]{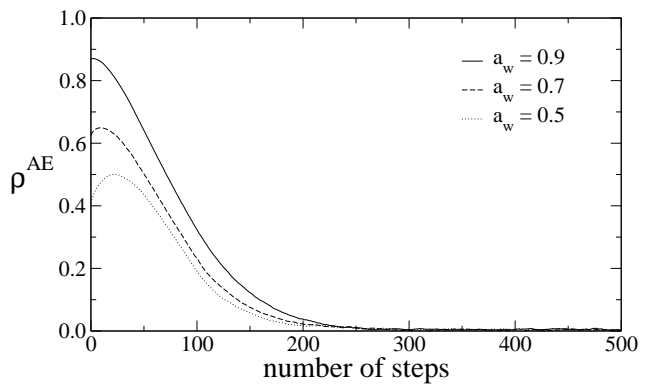}
  \caption{\label{fig:bg_overlap} The average overlap between student
    and generator as a function of the number of steps for $K=3$,
    $L=5$, and $N=100$.}
\end{figure}

Figures~\ref{fig:bg_pe} and~\ref{fig:bg_overlap} show that for $K=3$
hidden units, it is not possible to obtain an overlap to the
generating TPM by learning the sequence. Only if the initial overlap
between the generator and the student is very large there is a nonzero
probability $P_{sync}$ that the student will synchronize with the
generator. If it does not synchronize, the overlap between student and
generator decays to zero.

Summarizing, a TPM network generates a pseudorandom bit sequences
which cannot be predicted from part of the sequence. As a consequence,
for cryptographic applications, the TPM can be used to encrypt and
decrypt a secret message after it has generated a secret key.

\section{\label{sec:sync} Synchronization}

As shown in the preceding section, a TPM cannot learn the bit sequence
generated by another TPM since the two input vectors are completely
separated by the feedback mechanism. This also holds for
synchronization by mutual learning: With feedback, two networks cannot
be attracted to an identical time dependent state. Hence, to achieve
synchronization, we have to introduce an additional mechanism which
occasionally resets the two inputs to a common vector. This reset
occurs whenever the system has produced $R$ different output bits,
$\tau^A(t) \neq \tau^B(t)$. For $R=0$ we obtain synchronization
without feedback, which has been studied previously, and for large
values of $R$ the system does not synchronize. Accordingly, we have
added a new parameter in our algorithm which increases the
synchronization time as well as the difficulty to attack the system.
In the following two sections, we investigate synchronization and
security of the TPM with feedback quantitatively.

We consider two TPMs $A$ and $B$ which start with different random
weights and common random inputs. The feedback mechanism is defined as
follows.

\begin{enumerate}
  \renewcommand{\labelenumi}{(\roman{enumi})}
\item After each step $t$ the input is shifted,
  $x_{k,j}(t+1)=x_{k,j-1}(t)$ for $j>1$.
\item If the output bits agree, $\tau^A(t)=\tau^B(t)$, the output of
  each hidden unit is used as a new input bit,
  $x_{k,1}(t+1)=\sigma_k(t)$, otherwise all $K$ pairs of input bits
  $x_{k,1}(t)$ are set to common public random values.
\item After $R$ steps with different output, $\tau^A(t)\neq\tau^B(t)$,
  all input vectors are reset to public common random vectors,
  $x^A_{k,j}(t+1)=x^B_{k,j}(t+1)$.
\end{enumerate}

Feedback creates correlations between the weights and the inputs.
Therefore the system becomes sensitive to the learning rule. We find
that only for the anti-Hebbian rule, Eq.~(\ref{eq:anti-hebb}), the
components of the weights have a broad distribution. The entropy per
component is larger than 99\% of the maximal value $\ln(2L+1)$. For
the Hebbian or random walk rule, the entropy is much smaller, because
the values of the weights are pushed to the boundary values $\pm L$.
Therefore the network with the anti-Hebbian rule offers less
information to an attack than the two other rules.

\begin{figure}
  \centering
  \includegraphics[width=8.6cm]{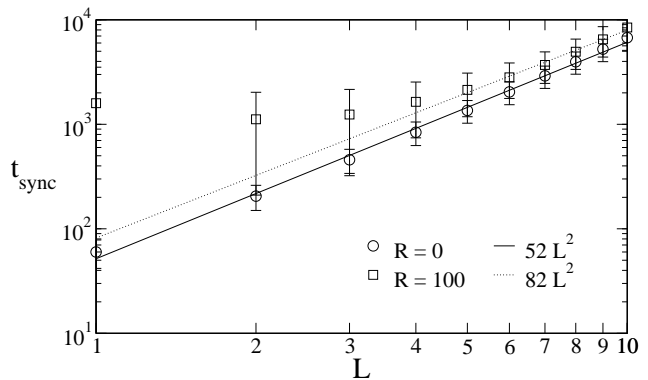}
  \caption{\label{fig:fb_time} Average synchronization time $t_{sync}$
    and its standard deviation as a function of $L$, found from 10 000
    simulation runs with $K=3$ and $N=10 000$. The line $52 L^2$ is a
    result of linear regression for $R=0$.}
\end{figure}

In Fig.~\ref{fig:fb_time} we have numerically calculated the average
synchronization time as a function of the number $L$ of components for
the anti-Hebbian rule. Obviously, there is a large deviation from the
scaling law $t_{sync} \propto L^2$ as observed for $R=0$. Our
simulations for larger values of $N$, which are not included here,
show that there exist strong finite size effects which do not allow to
derive a reliable scaling law from the numerical data.

\begin{figure}
  \centering
  \includegraphics[width=8.6cm]{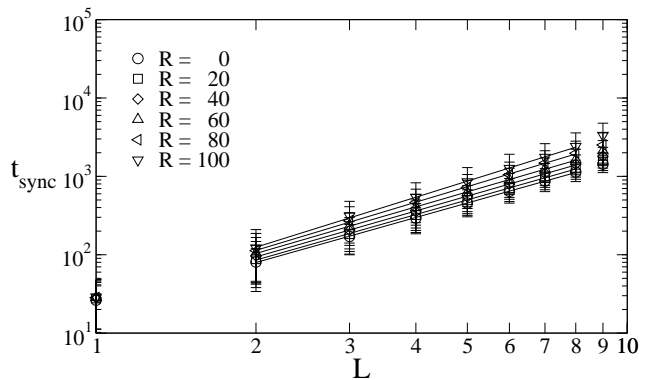}
  \caption{\label{fig:cfb_time} The synchronization time $t_{sync}$
    and its standard deviation as a function of $L$, averaged over
    10 000 runs of the iterative equations for $K=3$.}
\end{figure}

Fortunately, the limit $N \to \infty $ can be performed analytically.
The simulation of the $K N$ weights is replaced by a simulation of an
$(2L+1) \times (2L+1)$ overlap matrix $f_{a,b}^k$ for each hidden unit
$k$ which measures the fraction of weights which are in state $a$ for
the TPM $A$ and in state $b$ for $B$~\cite{Rosen-Zvi:2002:MLT,
  Rosen-Zvi:2002:CBN}.

We have extended this theory to the case of feedback. A new variable
$\lambda_k(t)$ is introduced which is defined as the fraction of input
components $x_{k,j}$ which are different between the corresponding
hidden units of $A$ and $B$. This variable changes with time, and it
influences the equation of motion for the overlap matrix
$f^k_{a,b}(t)$. Details are described in the Appendix.

Figure~\ref{fig:cfb_time} shows the results of this semianalytic
theory. Now, in the limit of $N \to \infty$, the average
synchronization time can be fitted to increase with a power of $L$,
roughly proportional to $L^2$. The data indicate that only the
prefactor but not the exponent depends on the strength $R$ of the
feedback; the prefactor seems to increase linearly with $R$.

Hence, if the network is large enough, feedback has only a small
effect on synchronization. In the following section we investigate the
effect of feedback on the security of the network: How does the
probability that an attacker is successful depend on the feedback
parameter $R$?

\section{Ensemble of attackers}

\begin{figure}
  \centering
  \includegraphics[width=8.6cm]{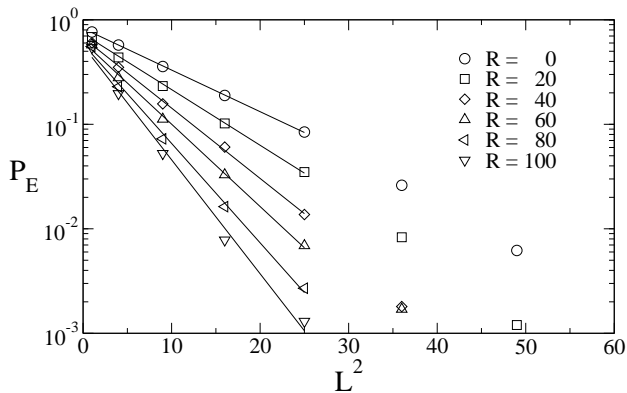}
  \caption{\label{fig:fb_pe} The success probability $P_E$ as a
    function of $L$, averaged over 10 000 simulations with $K=3$ and
    $N=1000$.}
\end{figure}

\begin{figure}
  \centering
  \includegraphics[width=8.6cm]{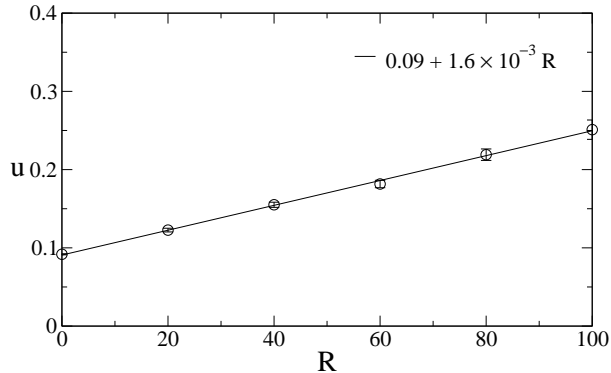}
  \caption{\label{fig:fb_u} The coefficient $u$ as a function of the
    feedback parameter $R$, calculated from the results shown in
    Fig.~\ref{fig:fb_pe}.}
\end{figure}

\begin{figure}
  \centering
  \includegraphics[width=8.6cm]{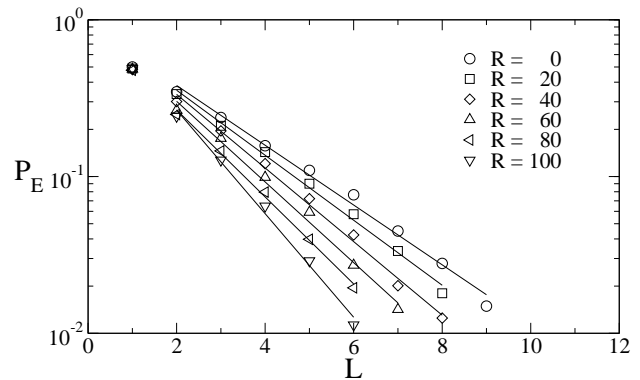}
  \caption{\label{fig:cfb_pe} The success probability $P_E$ as a
    function of $L$, found from 10 000 runs of the iterative equations
    for $K=3$.}
\end{figure}

\begin{figure}
  \centering
  \includegraphics[width=8.6cm]{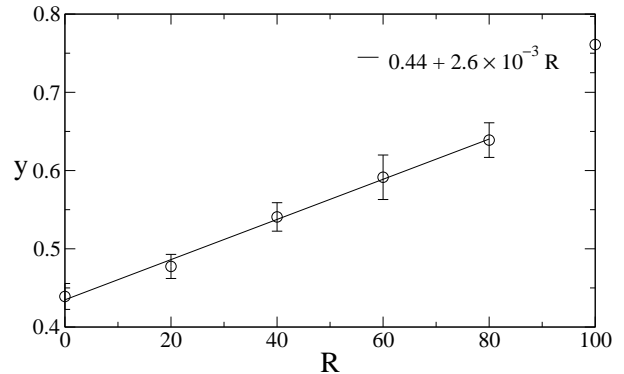}
  \caption{\label{fig:cfb_y} The coefficient $y$ as a function of the
    feedback parameter $R$, calculated from the results shown in
    Fig.~\ref{fig:cfb_pe}.}
\end{figure}

Up to now, the most successful attack on neural cryptography is the
geometric attack~\cite{Klimov:2002:ANC, Mislovaty:2002:SKE}. The
attacker $E$ uses the same TPM with an identical training step as the
two partners. That means, only for $\tau^A=\tau^B$ the attacker
performs a training step. When its output bit $\tau^E$ agrees with the
two partners, the attacker trains the hidden units which agree with
the common output. For $\tau^E \neq \tau^{A/B}$, however, the attacker
first inverts the output bit $\sigma_k$ for the hidden unit with the
smallest absolute value of the internal field and then performs the
usual training step.

For the geometric attack the probability $P_E$ that an attacker
synchronizes with $A$ and $B$ is nonzero. Consequently, if the
attacker uses an ensemble of sufficiently many networks there is a
good chance that at least one of them will find the secret key.

We have simulated an ensemble of attackers using the geometric attack
for the two TPMs with feedback and anti-Hebbian learning rule. Of
course, each attacking network uses the same feedback algorithm as the
two partner networks. Figure~\ref{fig:fb_pe} shows the results of our
numerical simulations. The success probability $P_E$ decreases with
the feedback parameter $R$. For the model parameters shown in
Fig.~\ref{fig:fb_pe} we find that $P_E$ can be fitted to an
exponential decrease with $L^2$,
\begin{equation}
  \label{eq:pe_u}
  P_E \propto e^{- u L^2}\,.
\end{equation}

The coefficient $u$ increases linearly with $R$, as shown in
Fig.~\ref{fig:fb_u}. The scaling [Eq.~(\ref{eq:pe_u})], however, is a
finite size effect. For large system sizes $N$, the success
probability decreases exponentially with $L$ instead of $L^2$,
\begin{equation}
  \label{eq:pe_y}
  P_E \propto e^{- y L}\,.
\end{equation}

This can be seen from the limit $N \to \infty$ which can be performed
with the analytic approach of the preceding section. Now the dynamics
of the system is described by a tensor $f_{a,b,e}^k$ for the three
networks $A$, $B$, and $E$ and corresponding variables
$\lambda_k^A,\lambda_k^B,\lambda_k^E$. Details are given in the
Appendix.

Figure~\ref{fig:cfb_pe} indicates the exponential scaling behavior
[Eq.~(\ref{eq:pe_y})] for several values of $R$. The coefficient $y$
increases linearly with $R$, as shown in Fig.~\ref{fig:cfb_y}.

These results show that feedback improves the security of neural
cryptography. The synchronization time, on the other side, increases,
too. Does the security of the system improve for constant effort of
the two partners?

\begin{figure}
  \centering
  \includegraphics[width=8.6cm]{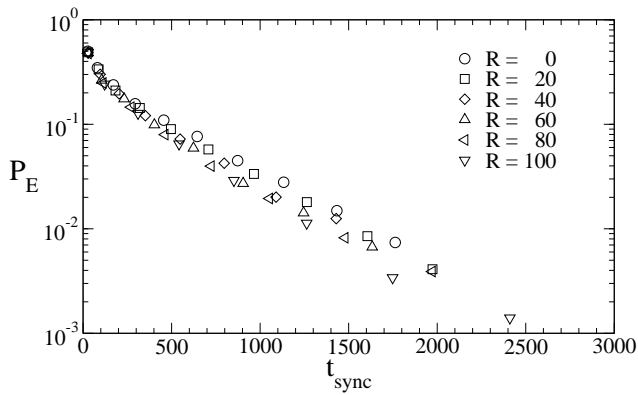}
  \caption{\label{fig:cfb_result} The success probability $P_E$ as a
    function of the average synchronization time $t_{sync}$,
    calculated from the results shown in Figs.~\ref{fig:cfb_time} and
    \ref{fig:cfb_pe}.}
\end{figure}

This question is answered in Fig.~\ref{fig:cfb_result} which shows the
probability $P_E$ as a function of the average synchronization time,
again for several values of the feedback parameter $R$. On the
logarithmic scale shown for $P_E$, the security does not depend much
on the feedback. For constant effort to find the secret key, feedback
yields a small improvement of security, only.

\section{Conclusions}

Neural cryptography is based on a delicate competition between
repulsive and attractive stochastic forces. A feedback mechanism has
been introduced which amplifies the repulsive part of these forces. We
find that feedback increases the synchronization time of two networks
and decreases the probability of a successful attack.

The numerical simulations up to $N=10^5$ do not allow to derive
reliable scaling laws, neither for the synchronization time nor for
the success probability. But the limit $N \to \infty$ which can be
performed analytically indicates that the scaling laws with respect to
the number $L$ of component values are not changed by the feedback,
only the respective coefficients are modified. The average
synchronization time increases with $L^2$ while the probability $P_E$
of a successful attack decreases exponentially with $L$, for huge
system sizes $N$.

Accordingly, the security of neural cryptography is improved by
including feedback in the training algorithm. But simultaneously the
effort to find the common key rises. We find that for a fixed
synchronization time, feedback yields a small improvement of security,
only.

After synchronization, the system is generating a pseudorandom bit
sequence which passed all tests on random numbers applied so far.
Even if another network is trained on this bit sequence it is not able
to extract some information on the statistical properties of the
sequence. Consequently, the neural cryptography cannot only generate a
secret key, but the same system can be used to encrypt and decrypt a
secret message, as well.

\appendix*
\section{Semianalytical calculation for synchronization with feedback}

In this appendix we describe our extension of the semianalytic
calculation~\cite{Rosen-Zvi:2002:MLT, Rosen-Zvi:2002:CBN} to the case
of feedback.

The effect of the feedback mechanism depends on the fraction $\Lambda$
of newly generated input elements $x_{k,j}$ per step and hidden unit.
In the numerical simulations presented in this paper $\Lambda$ is
equal to $N^{-1}$. In this case the effect of the feedback mechanism
vanishes in the limit $N \to \infty$. But it is also possible to
generate several input elements $x_{k,j}$ per hidden unit and step.
For that purpose one can multiply the output bit $\sigma_k$ with
$\Lambda N$ random numbers $z \in \{-1,+1\}$. As we want to compare
the results of the semianalytical approach with simulations for
$N=1000$, we set $\Lambda=10^{-3}$ in the following calculations.

In the case of two TPMs the development of the input noise $\lambda_k$
is given by
\begin{equation}
  \lambda_k^{+} = (1-\Lambda)\,\lambda_k +
  \Lambda\,\Theta(-\sigma_k^A\sigma_k^B)\,\Theta(\tau^A\tau^B)\,.
\end{equation}

At the beginning and after $R$ steps with $\tau^A \ne \tau^B$ all
variables $\lambda_k$ are set to zero (according to the algorithm
described in Sec.~\ref{sec:sync}).

The input noise generated by the feedback mechanism affects the output
of the hidden units. An input element with $x_{k,j}^B = -x_{k,j}^B$
causes the same output $\sigma_k^B$ as a change of sign in $w_{k,j}^B$
together with equal inputs for both $A$ and $B$. Therefore the
probability $\epsilon_{k,\mathrm{eff}}$ that two hidden units with
overlap $\rho_k$ and input error $\lambda_k$ disagree on the output
bit is given by
\begin{equation}
  \label{eq:eff}
  \epsilon_{k,\mathrm{eff}} = \frac{1}{\pi} \arccos
  (1-2\lambda_k)\rho_k\,.
\end{equation}

The distance $\epsilon_{k,\mathrm{eff}}$ between the hidden units of
$A$ and $B$ is used to choose the output bits $\sigma_k^A$ and
$\sigma_k^B$ with the correct probabilities in each
step~\cite{Rosen-Zvi:2002:MLT, Rosen-Zvi:2002:CBN}.

The feedback mechanism influences the equation of motion for the
overlap matrix $f^k_{a,b}$, too. Here we use additional variables
$\Delta_k^m = \Theta(\sigma_k^m \tau^m) \Theta(\tau^A \tau^B)$ to
determine if the weights of hidden unit $k$ in the TPM of $m \in
\{A,B,E\}$ change ($\Delta_k^m=1$) or not ($\Delta_k^m=0$). Therefore
we are able to describe the update of elements $f_{a,b}^k$ away from
the boundary ($-L<a,b<L$) in only one equation:
\begin{eqnarray}
  \label{eq:fnext}
  f^{k+}_{a,b} &=&
  \frac{1-\lambda_k}{2}
  (
    f^k_{a+\Delta_k^A,b+\Delta_k^B} +
    f^k_{a-\Delta_k^A,b-\Delta_k^B}
  )
  \nonumber \\ &+&
  \frac{1}{2} \lambda_k
  (
    f^k_{a+\Delta_k^A,b-\Delta_k^B} +
    f^k_{a-\Delta_k^A,b+\Delta_k^B}
  ).
\end{eqnarray}

The second term in Eq.~(\ref{eq:fnext}) which is proportional to
$\lambda_k$ shows the repulsive effect of the feedback mechanism.
Similar equations can be derived for elements on the boundary.

In the limit $N \to \infty$ the number of steps required to achieve
full synchronization diverges~\cite{Kinzel:2002:INN}. Because of that
one has to define a criterion which determines synchronization in
order to analyze the scaling of $t_{sync}$ using semianalytic
calculations. As in Ref.~\cite{Rosen-Zvi:2002:MLT} we choose the
synchronization criterion $\bar{\rho}^{AB} = \frac{1}{3} \sum_{k=1}^K
\rho^{AB}_k \geq 0.9$.

In order to analyze the geometric attack in the limit $N \to \infty$
one needs to extend the semianalytical calculation to three TPMs. In
this case the development of the input noise is given by the following
equations:
\begin{eqnarray}
  \lambda_k^{A+} &=& \Lambda\,\Theta(-\sigma_k^A\sigma_k^B)
  \,\Theta(-\sigma_k^A\sigma_k^E)\,\Theta(\tau^A\tau^B) \nonumber \\ 
  &+& (1-\Lambda)\,\lambda_k^A \,,
  \\
  \lambda_k^{B+} &=& \Lambda\,\Theta(-\sigma_k^B\sigma_k^A)
  \,\Theta(-\sigma_k^B\sigma_k^E)\,\Theta(\tau^A\tau^B) \nonumber \\
  &+& (1-\Lambda)\,\lambda_k^B \,,
  \\
  \lambda_k^{E+} &=& \Lambda\,\Theta(-\sigma_k^E\sigma_k^A)
  \,\Theta(-\sigma_k^E\sigma_k^B)\,\Theta(\tau^A\tau^B) \nonumber \\
  &+& (1-\Lambda)\,\lambda_k^E \,.
\end{eqnarray}

Analogical to Eq.~(\ref{eq:eff}) the distance
$\epsilon_{k,\mathrm{eff}}^{mn}$ between two hidden units can be
calculated from the overlap $\rho_k^{mn}$ and the variables
$\lambda_k^m$ and $\lambda_k^n$:
\begin{equation}
  \epsilon_{k,\mathrm{eff}}^{mn} = \frac{1}{\pi} \arccos
  (1-2\lambda_k^m-2\lambda_k^n) \rho_k^{mn}\,.
\end{equation}

But for the geometric attack the attacker $E$ needs to know the local
fields $h_k^E$. The joint probability distribution of $h_k^A$, $h_k^B$
and $h_k^E$ is given by~\cite{Rosen-Zvi:2002:MLT}
\begin{equation}
  P(h_k^A,h_k^B,h_k^E) = \frac{e^{-(1/2)(h_k^A , h_k^B , h_k^E)
      \mathcal{C}_k^{-1} (h_k^A , h_k^B , h_k^E)^T}}{\sqrt{(2\pi)^3
      \det \mathcal{C}_k}}\,.
\end{equation}
The covariance matrix in this equation describes the correlations
between the three neural networks:
\begin{equation}
  \mathcal{C}_k = \left(
    \begin{array}{ccc}
      Q_k^A                   &
      R_{k,\mathrm{eff}}^{AB} &
      R_{k,\mathrm{eff}}^{AE} \\
      R_{k,\mathrm{eff}}^{AB} &
      Q_k^B                   &
      R_{k,\mathrm{eff}}^{BE} \\
      R_{k,\mathrm{eff}}^{AE} &
      R_{k,\mathrm{eff}}^{BE} &
      Q_k^E
    \end{array}
  \right)\,.
\end{equation}
From the tensor $f^k_{a,b,e}$ and the variables $\lambda_k^m$ one can
easily calculate the elements of $\mathcal{C}_k$:
\begin{eqnarray}
    Q_k^A                                &=&
    \sum_{a,b,e=-L}^{L}{a^2 f^k_{a,b,e}} \,,\\
    Q_k^B                                &=&
    \sum_{a,b,e=-L}^{L}{b^2 f^k_{a,b,e}} \,,\\
    Q_k^E                                &=&
    \sum_{a,b,e=-L}^{L}{e^2 f^k_{a,b,e}} \,,\\
    R_{k,\mathrm{eff}}^{AB}              &=&
    (1-2\lambda_k^A-2\lambda_k^B)
    \sum_{a,b,e=-L}^{L}{ab f^k_{a,b,e}}  \,,\\
    R_{k,\mathrm{eff}}^{AE}              &=&
    (1-2\lambda_k^A-2\lambda_k^E)
    \sum_{a,b,e=-L}^{L}{ae f^k_{a,b,e}}  \,,\\
    R_{k,\mathrm{eff}}^{BE}              &=&
    (1-2\lambda_k^B-2\lambda_k^E)
    \sum_{a,b,e=-L}^{L}{be f^k_{a,b,e}}  \,.
\end{eqnarray}

We use a pseudorandom number generator to determine the values of
$h_k^A$, $h_k^B$, and $h_k^E$ in each step. The application of the
\emph{rejection method}~\cite{Hartmann:2002:OAP} ensures that the
local fields have the right joint probability distribution
$P(h_k^A,h_k^B,h_k^E)$. Then the output bits $\sigma_k^m$ of the
hidden units are given by $\sigma_k^m=\mathrm{sgn}(h_k^m)$. If $\tau^A
= \tau^B \ne \tau^E$ the hidden unit $k$ with the smallest absolute
local field $|h_k^E|$ is searched and its output $\sigma_k^E$ is
inverted (geometric attack). Afterwards the usual training of the
neural networks takes place.

The equation of motion for tensor elements $f_{a,b,e}^k$ away from the
boundary ($-L<a,b,e<L$) is given by
\begin{eqnarray}
  f^{k+}_{a,b,e} &=&
  \frac{1-\lambda_k^A-\lambda_k^B-\lambda_k^E}{2} \,
  f^k_{a+\Delta_k^A,b+\Delta_k^B,e+\Delta_k^E}
  \nonumber \\ &+&
  \frac{1-\lambda_k^A-\lambda_k^B-\lambda_k^E}{2} \,
  f^k_{a-\Delta_k^A,b-\Delta_k^B,e-\Delta_k^E}
  \nonumber \\ &+& 
  \frac{1}{2} \lambda_k^A \,
  f^k_{a-\Delta_k^A,b+\Delta_k^B,e+\Delta_k^E}
  \nonumber \\ &+&
  \frac{1}{2} \lambda_k^A \,
  f^k_{a+\Delta_k^A,b-\Delta_k^B,e-\Delta_k^E}
  \nonumber \\ &+&
  \frac{1}{2} \lambda_k^B \,
  f^k_{a+\Delta_k^A,b-\Delta_k^B,e+\Delta_k^E}
  \nonumber \\ &+&
  \frac{1}{2} \lambda_k^B \,
  f^k_{a-\Delta_k^A,b+\Delta_k^B,e-\Delta_k^E}
  \nonumber \\ &+&
  \frac{1}{2} \lambda_k^E \,
  f^k_{a+\Delta_k^A,b+\Delta_k^B,e-\Delta_k^E}
  \nonumber \\ &+&
  \frac{1}{2} \lambda_k^E \,
  f^k_{a-\Delta_k^A,b-\Delta_k^B,e+\Delta_k^E} \,.
\end{eqnarray}

Similar equations can be derived for elements on the boundary. An
attacker is considered successful if one of the conditions
$\bar{\rho}^{AE} \geq 0.9$ or $\bar{\rho}^{BE} \geq 0.9$ is achieved
earlier than the synchronization criterion $\bar{\rho}^{AB} \geq 0.9$.

\bibliography{paper}

\begin{thebibliography}{21}
\expandafter\ifx\csname natexlab\endcsname\relax\def\natexlab#1{#1}\fi
\expandafter\ifx\csname bibnamefont\endcsname\relax
  \def\bibnamefont#1{#1}\fi
\expandafter\ifx\csname bibfnamefont\endcsname\relax
  \def\bibfnamefont#1{#1}\fi
\expandafter\ifx\csname citenamefont\endcsname\relax
  \def\citenamefont#1{#1}\fi
\expandafter\ifx\csname url\endcsname\relax
  \def\url#1{\texttt{#1}}\fi
\expandafter\ifx\csname urlprefix\endcsname\relax\def\urlprefix{URL }\fi
\providecommand{\bibinfo}[2]{#2}
\providecommand{\eprint}[2][]{\url{#2}}

\bibitem[{\citenamefont{Hertz et~al.}(1991)\citenamefont{Hertz, Krogh, and
  Palmer}}]{Hertz:1991:ITN}
\bibinfo{author}{\bibfnamefont{J.}~\bibnamefont{Hertz}},
  \bibinfo{author}{\bibfnamefont{A.}~\bibnamefont{Krogh}}, \bibnamefont{and}
  \bibinfo{author}{\bibfnamefont{R.~G.} \bibnamefont{Palmer}},
  \emph{\bibinfo{title}{Introduction to the Theory of Neural Computation}}
  (\bibinfo{publisher}{Addison Wesley}, \bibinfo{address}{Redwood City},
  \bibinfo{year}{1991}).

\bibitem[{\citenamefont{Engel and Van~den Broeck}(2001)}]{Engel:2001:SML}
\bibinfo{author}{\bibfnamefont{A.}~\bibnamefont{Engel}} \bibnamefont{and}
  \bibinfo{author}{\bibfnamefont{C.}~\bibnamefont{Van~den Broeck}},
  \emph{\bibinfo{title}{Statistical Mechanics of Learning}}
  (\bibinfo{publisher}{Cambridge University Press},
  \bibinfo{address}{Cambridge}, \bibinfo{year}{2001}).

\bibitem[{\citenamefont{Metzler et~al.}(2000)\citenamefont{Metzler, Kinzel, and
  Kanter}}]{Metzler:2000:INN}
\bibinfo{author}{\bibfnamefont{R.}~\bibnamefont{Metzler}},
  \bibinfo{author}{\bibfnamefont{W.}~\bibnamefont{Kinzel}}, \bibnamefont{and}
  \bibinfo{author}{\bibfnamefont{I.}~\bibnamefont{Kanter}},
  \bibinfo{journal}{Phys. Rev. E} \textbf{\bibinfo{volume}{62}},
  \bibinfo{pages}{2555} (\bibinfo{year}{2000}).

\bibitem[{\citenamefont{Kinzel et~al.}(2000)\citenamefont{Kinzel, Metzler, and
  Kanter}}]{Kinzel:2000:DIN}
\bibinfo{author}{\bibfnamefont{W.}~\bibnamefont{Kinzel}},
  \bibinfo{author}{\bibfnamefont{R.}~\bibnamefont{Metzler}}, \bibnamefont{and}
  \bibinfo{author}{\bibfnamefont{I.}~\bibnamefont{Kanter}},
  \bibinfo{journal}{J. Phys. A} \textbf{\bibinfo{volume}{33}},
  \bibinfo{pages}{L141} (\bibinfo{year}{2000}).

\bibitem[{\citenamefont{Kanter et~al.}(2002)\citenamefont{Kanter, Kinzel, and
  Kanter}}]{Kanter:2002:SEI}
\bibinfo{author}{\bibfnamefont{I.}~\bibnamefont{Kanter}},
  \bibinfo{author}{\bibfnamefont{W.}~\bibnamefont{Kinzel}}, \bibnamefont{and}
  \bibinfo{author}{\bibfnamefont{E.}~\bibnamefont{Kanter}},
  \bibinfo{journal}{Europhys. Lett.} \textbf{\bibinfo{volume}{57}},
  \bibinfo{pages}{141} (\bibinfo{year}{2002}).

\bibitem[{\citenamefont{Kinzel and Kanter}(2003)}]{Kinzel:2003:DGI}
\bibinfo{author}{\bibfnamefont{W.}~\bibnamefont{Kinzel}} \bibnamefont{and}
  \bibinfo{author}{\bibfnamefont{I.}~\bibnamefont{Kanter}},
  \bibinfo{journal}{J. Phys. A} \textbf{\bibinfo{volume}{36}},
  \bibinfo{pages}{11 173} (\bibinfo{year}{2003}).

\bibitem[{\citenamefont{Stinson}(1995)}]{Stinson:1995:CTP}
\bibinfo{author}{\bibfnamefont{D.~R.} \bibnamefont{Stinson}},
  \emph{\bibinfo{title}{Cryptography: Theory and Practice}}
  (\bibinfo{publisher}{CRC Press}, \bibinfo{address}{Boca Rotan, FL},
  \bibinfo{year}{1995}).

\bibitem[{\citenamefont{Klimov et~al.}(2003)\citenamefont{Klimov, Mityagin, and
  Shamir}}]{Klimov:2002:ANC}
\bibinfo{author}{\bibfnamefont{A.}~\bibnamefont{Klimov}},
  \bibinfo{author}{\bibfnamefont{A.}~\bibnamefont{Mityagin}}, \bibnamefont{and}
  \bibinfo{author}{\bibfnamefont{A.}~\bibnamefont{Shamir}}, in
  \emph{\bibinfo{booktitle}{Advances in Cryptology - ASIACRYPT 2002}}, edited
  by \bibinfo{editor}{\bibfnamefont{Y.}~\bibnamefont{Zheng}}
  (\bibinfo{publisher}{Springer}, \bibinfo{address}{Heidelberg},
  \bibinfo{year}{2003}), p. \bibinfo{pages}{288}.

\bibitem[{\citenamefont{Kinzel and
  Kanter}(2002{\natexlab{a}})}]{Kinzel:2002:INN}
\bibinfo{author}{\bibfnamefont{W.}~\bibnamefont{Kinzel}} \bibnamefont{and}
  \bibinfo{author}{\bibfnamefont{I.}~\bibnamefont{Kanter}}, in
  \emph{\bibinfo{booktitle}{Advances in Solid State Physics}}, edited by
  \bibinfo{editor}{\bibfnamefont{B.}~\bibnamefont{Kramer}}
  (\bibinfo{publisher}{Springer}, \bibinfo{address}{Berlin},
  \bibinfo{year}{2002}{\natexlab{a}}), vol.~\bibinfo{volume}{42}, pp.
  \bibinfo{pages}{383--391}.

\bibitem[{\citenamefont{Kinzel and
  Kanter}(2002{\natexlab{b}})}]{Kinzel:2002:NC}
\bibinfo{author}{\bibfnamefont{W.}~\bibnamefont{Kinzel}} \bibnamefont{and}
  \bibinfo{author}{\bibfnamefont{I.}~\bibnamefont{Kanter}}
  (\bibinfo{year}{2002}{\natexlab{b}}), \eprint{cond-mat/0208453}.

\bibitem[{\citenamefont{Mislovaty et~al.}(2002)\citenamefont{Mislovaty,
  Perchenok, Kinzel, and Kanter}}]{Mislovaty:2002:SKE}
\bibinfo{author}{\bibfnamefont{R.}~\bibnamefont{Mislovaty}},
  \bibinfo{author}{\bibfnamefont{Y.}~\bibnamefont{Perchenok}},
  \bibinfo{author}{\bibfnamefont{W.}~\bibnamefont{Kinzel}}, \bibnamefont{and}
  \bibinfo{author}{\bibfnamefont{I.}~\bibnamefont{Kanter}},
  \bibinfo{journal}{Phys. Rev E} \textbf{\bibinfo{volume}{66}},
  \bibinfo{pages}{066102} (\bibinfo{year}{2002}).

\bibitem[{\citenamefont{Kanter and Kinzel}(2003)}]{Kanter:2002:TNN}
\bibinfo{author}{\bibfnamefont{I.}~\bibnamefont{Kanter}} \bibnamefont{and}
  \bibinfo{author}{\bibfnamefont{W.}~\bibnamefont{Kinzel}}, in
  \emph{\bibinfo{booktitle}{Proceedings of the XXII Solvay Conference on
  Physics on the Physics of Communication}}, edited by
  \bibinfo{editor}{\bibfnamefont{I.}~\bibnamefont{Antoniou}},
  \bibinfo{editor}{\bibfnamefont{V.~A.} \bibnamefont{Sadovnichy}},
  \bibnamefont{and} \bibinfo{editor}{\bibfnamefont{H.}~\bibnamefont{Wather}}
  (\bibinfo{publisher}{World Scientific}, \bibinfo{address}{Singapore},
  \bibinfo{year}{2003}), p. \bibinfo{pages}{631}.

\bibitem[{\citenamefont{Rosen-Zvi
  et~al.}(2002{\natexlab{a}})\citenamefont{Rosen-Zvi, Klein, Kanter, and
  Kinzel}}]{Rosen-Zvi:2002:MLT}
\bibinfo{author}{\bibfnamefont{M.}~\bibnamefont{Rosen-Zvi}},
  \bibinfo{author}{\bibfnamefont{E.}~\bibnamefont{Klein}},
  \bibinfo{author}{\bibfnamefont{I.}~\bibnamefont{Kanter}}, \bibnamefont{and}
  \bibinfo{author}{\bibfnamefont{W.}~\bibnamefont{Kinzel}},
  \bibinfo{journal}{Phys. Rev. E} \textbf{\bibinfo{volume}{66}},
  \bibinfo{pages}{066135} (\bibinfo{year}{2002}{\natexlab{a}}).

\bibitem[{com()}]{comment}
\bibinfo{note}{A similar mechanism for a noise which increases the security of
  the network was recently found by combining synchronization of neural
  networks and chaotic maps. See Ref. \cite{Mislovaty:2003:PCC}}.

\bibitem[{\citenamefont{Mislovaty et~al.}(2003)\citenamefont{Mislovaty, Klein,
  Kanter, and Kinzel}}]{Mislovaty:2003:PCC}
\bibinfo{author}{\bibfnamefont{R.}~\bibnamefont{Mislovaty}},
  \bibinfo{author}{\bibfnamefont{E.}~\bibnamefont{Klein}},
  \bibinfo{author}{\bibfnamefont{I.}~\bibnamefont{Kanter}}, \bibnamefont{and}
  \bibinfo{author}{\bibfnamefont{W.}~\bibnamefont{Kinzel}},
  \bibinfo{journal}{Phys. Rev. Lett.} \textbf{\bibinfo{volume}{91}},
  \bibinfo{pages}{118701} (\bibinfo{year}{2003}).

\bibitem[{\citenamefont{Eisenstein et~al.}(1995)\citenamefont{Eisenstein,
  Kanter, Kessler, and Kinzel}}]{Eisenstein:1995:GPT}
\bibinfo{author}{\bibfnamefont{E.}~\bibnamefont{Eisenstein}},
  \bibinfo{author}{\bibfnamefont{I.}~\bibnamefont{Kanter}},
  \bibinfo{author}{\bibfnamefont{D.}~\bibnamefont{Kessler}}, \bibnamefont{and}
  \bibinfo{author}{\bibfnamefont{W.}~\bibnamefont{Kinzel}},
  \bibinfo{journal}{Phys. Rev. Lett.} \textbf{\bibinfo{volume}{74}},
  \bibinfo{pages}{6} (\bibinfo{year}{1995}).

\bibitem[{\citenamefont{Zhu and Kinzel}(1998)}]{Zhu:1998:APS}
\bibinfo{author}{\bibfnamefont{H.}~\bibnamefont{Zhu}} \bibnamefont{and}
  \bibinfo{author}{\bibfnamefont{W.}~\bibnamefont{Kinzel}},
  \bibinfo{journal}{Neural Comput.} \textbf{\bibinfo{volume}{10}},
  \bibinfo{pages}{2219} (\bibinfo{year}{1998}).

\bibitem[{\citenamefont{Metzler et~al.}(2001)\citenamefont{Metzler, Kinzel,
  Ein-Dor, and Kanter}}]{Metzler:2001:GUT}
\bibinfo{author}{\bibfnamefont{R.}~\bibnamefont{Metzler}},
  \bibinfo{author}{\bibfnamefont{W.}~\bibnamefont{Kinzel}},
  \bibinfo{author}{\bibfnamefont{L.}~\bibnamefont{Ein-Dor}}, \bibnamefont{and}
  \bibinfo{author}{\bibfnamefont{I.}~\bibnamefont{Kanter}},
  \bibinfo{journal}{Phys. Rev. E} \textbf{\bibinfo{volume}{63}},
  \bibinfo{pages}{056126} (\bibinfo{year}{2001}).

\bibitem[{\citenamefont{Knuth}(1981)}]{Knuth:1981:SA}
\bibinfo{author}{\bibfnamefont{D.~E.} \bibnamefont{Knuth}},
  \emph{\bibinfo{title}{Seminumerical Algorithms}}, vol.~\bibinfo{volume}{2} of
  \emph{\bibinfo{series}{The Art of Computer Programming}}
  (\bibinfo{publisher}{Addison-Wesley}, \bibinfo{address}{Redwood City},
  \bibinfo{year}{1981}).

\bibitem[{\citenamefont{Rosen-Zvi
  et~al.}(2002{\natexlab{b}})\citenamefont{Rosen-Zvi, Kanter, and
  Kinzel}}]{Rosen-Zvi:2002:CBN}
\bibinfo{author}{\bibfnamefont{M.}~\bibnamefont{Rosen-Zvi}},
  \bibinfo{author}{\bibfnamefont{I.}~\bibnamefont{Kanter}}, \bibnamefont{and}
  \bibinfo{author}{\bibfnamefont{W.}~\bibnamefont{Kinzel}},
  \bibinfo{journal}{J. Phys. A} \textbf{\bibinfo{volume}{35}},
  \bibinfo{pages}{L707} (\bibinfo{year}{2002}{\natexlab{b}}).

\bibitem[{\citenamefont{Hartmann and Rieger}(2002)}]{Hartmann:2002:OAP}
\bibinfo{author}{\bibfnamefont{A.~K.} \bibnamefont{Hartmann}} \bibnamefont{and}
  \bibinfo{author}{\bibfnamefont{H.}~\bibnamefont{Rieger}},
  \emph{\bibinfo{title}{Optimization Algorithms in Physics}}
  (\bibinfo{publisher}{Wiley-VCH}, \bibinfo{address}{Berlin},
  \bibinfo{year}{2002}).

\end{thebibliography}

\end{document}